\documentclass[pra,twocolumn,showpacs]{revtex4-1}
\usepackage{dcolumn}
\usepackage{bm}
\usepackage{latexsym}
\usepackage{amsfonts}
\usepackage{amsmath}
\usepackage{amsthm}
\usepackage{amssymb}
\usepackage{amsfonts}
\usepackage{bm}
\usepackage{bbm}
\usepackage{hyperref}
\usepackage[dvips]{graphicx}
\usepackage[usenames]{color}
\usepackage{makeidx}

\newcommand{\half}{\frac{1}{2}}
\newcommand{\be}{\begin{equation}}
\newcommand{\ee}{\end{equation}}
\newcommand{\bee}{\begin{equation*}}
\newcommand{\eee}{\end{equation*}}
\newcommand{\bq}{\bm{q}}
\newcommand{\dq}{\text{d}\bm{q}}
\newcommand{\dk}{\text{d}\bm{k}}
\newcommand{\dz}{\mathrm{d}z}

\newcommand{\dbk}{\mathrm{d}\bk}
\newcommand{\dbq}{\mathrm{d}\bq}
\newcommand{\bk}{\bm{k}}

\newcommand{\epair}{\varepsilon_{\text{pair}}^{}}
\newcommand{\da}{\dagger}

\newcommand{\br}{\bm{r}}
\newcommand{\eps}[1]{\varepsilon_{#1}^{}}
\newcommand{\om}{\omega}

\newcommand{\kf}{k_F^{}}
\newcommand{\kb}{k_B^{}}
\newcommand{\eb}{\varepsilon_{B}^{}}
\newcommand{\cg}{\mathcal{G}}
\newcommand{\tmat}{T^{}_{\uparrow\downarrow}}
\newcommand{\dar}{\downarrow}
\newcommand{\ra}{\rangle}
\newcommand{\la}{\langle}
\newcommand{\uar}{\uparrow}
\newcommand{\mud}{\mu_{\uparrow}}
\newcommand{\mua}{\mu_{\downarrow}}
\newcommand{\nf}{n_{F}^{}}
\newcommand{\nb}{n_{B}^{}}
\newcommand{\re}{\text{Re}}
\newcommand{\im}{\text{Im}}

\newcommand{\asc}{a_{2D}^{}}
\newcommand{\ass}{a_{3D}^{}}
\newcommand{\omrf}{\Omega_{\mathrm{rf}}^{}}
\newcommand{\omrff}{\Omega_{\mathrm{rf}}^{2}}
\newcommand{\irf}{I_{\mathrm{rf}}^{}}
\newcommand{\oml}{\omega_{\mathrm{rf}}^{}}

\begin{document}

\title{Pairing and radio-frequency spectroscopy in two-dimensional Fermi gases}
\author{Ville Pietil\"a\vspace{3pt}}
\address{Physics Department, Harvard University, Cambridge, Massachusetts 02138, USA
\vspace{3pt}}
\date{\today}
\begin{abstract}
We theoretically study the normal phase properties of strongly interacting two-component Fermi 
gases in two spatial dimensions. In the limit of weak attraction, we find that the gas 
can be described in terms of effective polarons. As the attraction between fermions increases, we 
find a crossover from a gas of non-interacting polarons to a pseudogap regime. We investigate 
how this crossover is manifested in the radio-frequency (rf) spectroscopy. Our findings qualitatively 
explain the differences in the recent  rf spectroscopy measurements of two-dimensional 
Fermi gases [Sommer {\it et al.},  Phys.~Rev.~Lett.~{\bf 108}, 045302 (2012) and Zhang {\it et al.}, 
Phys.~Rev.~Lett.~{\bf 108}, 235302 (2012)].
\end{abstract}

\maketitle

\section{Introduction}

Pairing of fermions in two spatial dimensions has become one of the main themes in 
condensed-matter physics due to the discovery of high-temperature superconductivity in copper 
oxide compounds~\cite{Abrahams:2010}. Ultracold gases of fermionic atoms provide a flexible 
platform for testing the various pairing mechanisms. Advances in manipulating atoms 
in optical lattices can ultimately lead to direct simulation of materials relevant to 
high-temperature superconductivity. To achieve this milestone, one first needs to  
understand pairing in simpler systems, such as Fermi gases confined to continuous 
two-dimensional (2D) geometries. Several recent experiments have probed different aspects of 
strongly interacting 2D Fermi gases, ranging from polaron physics~\cite{Frohlich:2011, 
Koschorreck:2012, Zhang:2012} to pairing and confinement-induced 
molecules~\cite{Dyke:2011,Sommer:2012,Feld:2011,Baur:2012}. Another interesting direction is 
a dimensional crossover from two dimensions to three dimensions~\cite{Dyke:2011, 
Sommer:2012}.

One of the intriguing aspects of both high-temperature superconductors and atomic Fermi gases 
is the possible existence of preformed Cooper pairs above the superfluid transition 
temperature~\cite{Abrahams:2010,Loktev:2001,Perali:2002,Rohe:2001,Hu:2010,
Tsuchiya:2009,Chien:2010}. This regime is referred to as the pseudogap phase, and in the 
context of Fermi gases, the properties of the pseudogap regime have been experimentally probed 
in both two-dimensional and three-dimensional (3D) systems~\cite{Gaeblar:2010,
Perali:2011,Feld:2011,Stewart:2008}. The nature of the normal state of a strongly interacting Fermi 
gas is still an open question since it can also be interpreted in terms of the Fermi-liquid 
framework~\cite{Nascimbene:2011}. Furthermore, recent experiments have suggested that 
a 2D Fermi gas can be described as a gas of noninteracting polarons even in the absence of any 
population imbalance~\cite{Zhang:2012}, while other experiments indicate the existence of 
confinement-induced molecules in the normal phase~\cite{Sommer:2012}.

In this work, we probe the properties of the normal state of strongly interacting 2D Fermi gases. In 
particular, we aim to qualitatively explain the differences in the recent experiments performed in 
the Zwierlein group at Massachusetts Institute of Technology (MIT)~\cite{Sommer:2012} and in 
the Thomas group at North Carolina State University (NCSU)~\cite{Zhang:2012}. We find that the 
normal state of weakly attractive Fermi gases can be described in terms of effective polarons,  
whereas strongly attractive Fermi gases are characterized by a Bardeen-Cooper-Schrieffer 
(BCS) -like effective dispersion and suppression of a single-particle density of states (DOS) 
near the Fermi energy. Such characteristics are commonly attributed to the pseudogap 
regime~\cite{Perali:2002, Stewart:2008,Tsuchiya:2009,Haussmann:2009,Hu:2010, 
Gaeblar:2010,Mueller:2011}. Our analysis suggests a crossover from a gas of non-interacting 
polarons to the pseudogap regime with increasing attraction and qualitatively explains the 
different experimental results reported in Refs.~\cite{Sommer:2012,Zhang:2012}. 
We note that the earlier experiments reported in Ref.~\cite{Frohlich:2011} regarding  molecule 
formation in 2D~Fermi gases have already been interpreted in terms of dynamically created 
polarons~\cite{Schmidt:2011} and therefore we do not discuss this experiment in the present 
work.

The experiments reported in Refs.~\cite{Sommer:2012,Zhang:2012} used   
radio-frequency (rf) spectroscopy to probe the properties of 2D Fermi gases.
Both experiments can be schematically described by an interacting, population 
balanced initial mixture of spin-$\uparrow$ and spin-$\downarrow$ atoms. A short rf pulse is 
applied to convert spin-$\downarrow$ atoms into a final-state $|f\ra$ and the number of 
converted atoms (or, equivalently, atoms remaining in state $|\downarrow\ra$) is subsequently 
measured. By varying the detuning of the rf pulse from the bare atomic transition, one obtains 
information on the single-particle excitation spectrum. For this reason, rf spectroscopy has been 
extensively used to probe the pairing mechanisms in strongly interacting Fermi 
gases~\cite{Chin:2004,Shin:2007,Schunck:2007,Schunck:2008,Schirotzek:2008,Stewart:2008}. 
Although both experiments probed, in principle, the same initial system, the 
final conclusions were quite different. The MIT experiment~\cite{Sommer:2012} was 
interpreted in terms of confinement-enhanced pairing~\cite{Randeria:1989}, whereas the NCSU 
experiment~\cite{Zhang:2012} found the resonances in the rf spectra to be best explained by 
transitions between polarons in the initial and final-states. 

To analyze the different contributions to the experimentally measured rf spectra, 
one needs to take into account interactions between the initial-state atoms ($ |\uparrow\ra$ and 
$|\downarrow\ra$) and the final-state atoms ($ |\uparrow\ra$ and $ |f\ra$). On the other hand, 
interactions between atoms in states $|\downarrow\ra$ and $|f\ra$ are irrelevant since the rf pulse 
coherently rotates the atomic spin~\cite{Stoof:2009}. We note that both the MIT and the NCSU 
experiments correspond to relatively weak final-state 
interactions~\cite{Sommer:2012,Zhang:2012}. Therefore we consider the final-state interactions 
only phenomenologically and concentrate on the strong initial-state interactions. 
The contribution of the final-state interactions to the MIT experiment has recently been discussed 
in Ref.~\cite{Langmack:2012}. In Sec.~\ref{t_matrix}, we describe the non-self-consistent 
$T$-matrix formalism which we use to calculate the spectral properties of the initial- and the 
final-state atoms. Section~\ref{chem_pot} describes the different schemes for calculating the 
chemical potential of the population balanced initial state.  The quasiparticle spectrum of the 
initial- and the final-state atoms is discussed in Sec.~\ref{spec_properties}. The rf spectroscopy 
corresponding to the experiments in Refs.~\cite{Sommer:2012,Zhang:2012} is studied in 
Sec.~\ref{rf}, and concluding remarks are presented in Sec.~\ref{discussion}.


\section{\label{t_matrix}T-matrix and the ladder approximation}

\subsection{Initial state}

\begin{figure}[h!]
\begin{center}
\includegraphics[width=0.3\textwidth]{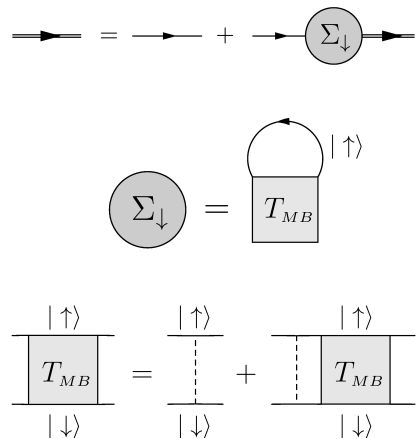}
\end{center}
\caption{\label{diagrams} Ladder approximation for the self-energy. We restrict the analysis 
to the fluctuation part since the exchange part is important only in the case of magnetic ordering. 
Furthermore, we assume equal densities of spin-$\downarrow$ and spin-$\uparrow$ atoms, 
which makes the diagrams interchangeable with respect to spin-$\uparrow$ and spin-$\downarrow$.}
\end{figure}

We study the properties of the population balanced initial state using a non-self-consistent 
$T$-matrix approximation where the $T$-matrix is computed by summing over all ladder 
diagrams. This approximation has been extensively utilized in the literature to study the 
properties of Fermi gases in the normal phase~\cite{Nozieres:1985,SchmittRink:1989, 
Tokumitu:1993,Tsuchiya:2009, Rohe:2001,Perali:2002,Punk:2007}. Alternative theoretical 
approaches have been discussed in Ref.~\cite{Chien:2010}. The dressed Green's function can 
be calculated from the Dyson's equation,
\begin{equation}
\label{dyson}
\cg_{\sigma}^{-1}(\bk,i\om_n) = \cg_{0,\sigma}^{-1}(\bk,i\om_n) - \Sigma_{\sigma}(\bk,i\om_n),
\end{equation}
where the bare Green's function is given by 
$\cg_{0,\sigma}^{-1}(\bk,i\om_n) = i\om_n - \eps{\bk} + \mu_{\sigma}$,
and $\sigma = \uar,\dar$. The fermionic Matsubara frequencies are defined as 
$\om_n^{} = (2n +1)\pi/\beta$ and $\beta = 1/\kb T$. Furthermore, we have set $\hbar = 1$.
Within the non-self-consistent ladder approximation~\cite{Nozieres:1985,SchmittRink:1989, 
Tokumitu:1993,Tsuchiya:2009, Rohe:2001,Perali:2002,Punk:2007}, the self-energy is given by 
the $T$-matrix and the bare Green's function, 
\begin{align}
\Sigma_{\sigma}(\bk,i\omega_n^{}) = \frac{1}{\beta}\sum_{i\Omega_m^{}} & \int 
\frac{\dbq}{(2\pi)^2}\,\tmat(\bq,i\Omega_m^{})\,\times \notag \\
\label{self-energy}
& \quad \quad \mathcal{G}_{0,-\sigma}^{}(\bq-\bk,i\Omega_m^{}-
i\omega_n^{}).
\end{align}
Notation $-\sigma$ indicates a spin opposite of $\sigma$ and $\Omega_n = 2\pi n/\beta$ denote 
the bosonic Matsubara frequencies. The many-body $T$-matrix can be 
calculated from the Bethe-Salpeter equation using the ladder approximation and the bare 
Green's function
\begin{equation}
\frac{1}{\tmat(\bq,i\Omega_m)} = \frac{1}{V_0} + \chi_0^{}(\bq,i\Omega_m),
\end{equation}
such that the polarization operator is given by 
\begin{align}
\label{gen_chi}
\chi_0^{}(\bq,i\Omega_m) = \frac{1}{\beta}\sum_{i\om_n}\int\frac{\dbk}{(2\pi)^2}\, 
 \cg_{0,\uar}(\bq & -\bk, i\Omega_m-i\omega_n) \notag \\ 
& \times \, \cg_{0,\dar}(\bk,i\omega_n).
\end{align}
The ladder approximation for computing the self-energy is illustrated in 
Fig.~\ref{diagrams}.

We assume the gas has equal densities of spin-$\downarrow$ and 
spin-$\uparrow$ particles, which implies $\mu_\dar=\mu_\uar = \mu$. To cancel the ultraviolet 
(UV) divergence associated with the polarization operator, we use the vacuum $T$-matrix 
\begin{equation}
\label{t0}
T_0(\om) = \frac{4\pi/m}{\ln(\eb/\om) + i\pi},
\end{equation}  
where $\eb = 1/ma_{2D}^2 > 0$ is the two-body bound-state energy. The many-body 
$T$-matrix takes the form
\be
\frac{1}{\tmat(\bq,i\Omega_m)} = \frac{1}{T_0^{}(i\Omega_m + 2\mu-\half\eps{\bq})} + 
\chi_0^{\text{reg}}(\bq,i\Omega_m),
\ee 
where the regularized polarization operator $\chi_0^{\text{reg}}(\bq,i\Omega_m)$  is given by 
\begin{align}
\label{chi_reg}
\chi_0^{\text{reg}} & (\bq,i\Omega_m) =  \notag \\
&\int\frac{\dbk}{(2\pi)^2}\,\frac{\nf(\eps{\bk+\bq/2} - \mu) + 
\nf(\eps{\bk-\bq/2}-\mu)}{i\Omega_m-2\eps{\bk} - \frac{1}{2}\eps{\bq}+2\mu},
\end{align}
and $\nf(\om)$ denotes the Fermi function. In this work, we are interested in finite temperatures 
above the superfluid phase-transition temperature, and the polarization operator in 
Eq.~\eqref{chi_reg} has to be computed numerically.

After the analytic continuation $i\om_n^{} \rightarrow \om + i0^+_{}$, the retarded self-energy 
$\Sigma_{\downarrow}^{}(\bq,\om+i0^+_{})$ can be computed from Eq.~\eqref{self-energy}  using 
contour integration. Since the $T$-matrix may have poles away from the real axis, we cannot 
directly apply the spectral representation of the $T$-matrix to compute the self-energy. However, 
assuming that the $T$-matrix has poles either on the real axis 
or symmetrically with respect to the real axis, we arrive at a simple result for the imaginary part 
of the self-energy,
\begin{align}
\label{im_se}
\im\,\Sigma_{\downarrow}^{}(\bq,\om) = \int & \frac{\dk}{(2\pi)^2}  \big[\nb(\om+\eps{\bk}-\mu)+
\nf(\eps{\bk}-\mu) \big] \notag \\ & \quad \times \,\im\,\tmat(\bq+\bk,\om+\eps{\bk}-\mu),
\end{align}
where $\nb(\om)$ denotes the Bose distribution. We have numerically verified that, for the 
parameter regime considered in this work, the $T$-matrix has poles at most on the real axis. 
The real part of the self-energy is calculated using the Kramers-Kronig relation 
\be
\label{kk_rel}
\re\,\Sigma_{\downarrow}^{}(\bq,\om) = \mathcal{P}\,\int_{-\infty}^{\infty}\frac{\text{d}z}{\pi}\,
\frac{\im\,\Sigma_{\downarrow}^{}(\bq,z)}{z-\om},
\ee
where $\mathcal{P}$ denotes principal-value integration.

In 3D systems the appearance of poles in the $T$-matrix with a nonzero imaginary part is 
associated with the onset of the superfluid phase. This is the Thouless 
criterion~\cite{Thouless:1960}, and it can be used to identify the superfluid transition 
temperature $T_c^{}$. In 2D Fermi gases, the superfluid transition takes 
place via the Berezinskii-Kosterlitz-Thouless (BKT) transition, which is not captured by our 
non-self-consistent  $T$-matrix theory~\cite{Loktev:2001}. In order to access the BKT physics, 
one needs to either consider the vertex corrections to the $T$-matrix formalism~\cite{Loktev:2001} 
or explicitly include phase fluctuations to a mean-field formalism~\cite{Botelho:2006}. 
Our non-self-consistent $T$-matrix theory should therefore be 
used at temperatures above $T_c^{}$ where it provides a reasonable description of the 
system. 

The retarded Green's function is obtained by analytic continuation $G_{\downarrow}^{R}(\bq,\om) = 
\mathcal{G}_{\downarrow}^{}(\bq,i\om_n\rightarrow\om+i0^+)$ from the thermal Green's function 
given by the Dyson's equation~\eqref{dyson}. The corresponding spectral function is defined as 
$A_{\downarrow}^{}(\bq,\om) = -2\,\im\,G_{\downarrow}^{R}(\bq,\om)$, and it satisfies a sum rule
$\int_{-\infty}^{\infty}\frac{\mathrm{d}\om}{2\pi}\,A_{\downarrow}^{}(\bq,\om) = 1$. 
We use this sum rule to verify the consistency of the numerical calculation.
To obtain a physically correct spectral function for a strongly interacting initial gas of 
spin-$\uparrow$ and spin-$\downarrow$ atoms, the chemical potential $\mu$ has to be 
determined self-consistently. In the next section, we discuss the different self-consistent schemes 
for computing $\mu$.

\subsection{Final state}

In Ref.~\cite{Zhang:2012}, the experimental findings were interpreted in terms of transitions 
between initial- and final-state polarons. In order to test this hypothesis, we consider the 
properties of the final-state atoms when they are dressed by the interactions with the initial 
state atoms. We take the interactions between atoms in hyperfine spin states 
$|\uparrow\ra$ and $| f\ra$ into account, using again the $T$-matrix formalism~\cite{Schmidt:2011}. 
For the final-state atoms, the regularized polarization operator is given by 
\begin{align}
\label{chi_final}
\chi_{0}^{\text{reg}} & (\bq,  i\Omega_m)  = \notag \\
&\int\frac{\dbk}{(2\pi)^2}\,\frac{\nf(\eps{\bk+\bq/2} - \mu) + 
\nf(\eps{\bk-\bq/2}+\Delta_{\uparrow f}) }
{i\Omega_m-2\eps{\bk} - \frac{1}{2}\eps{\bq}-\Delta_{\uparrow f}+\mu},
\end{align}
where $\Delta_{\uparrow f}$ is the level splitting between states $|\uparrow\ra$ and 
$| f\ra$. We have also denoted $\mu = \mua$. Since the final-state is initially empty, we have taken 
the final-state chemical potential to be $\mu_f^{}=0$. We use the vacuum $T$-matrix to 
regularize the UV divergence. This introduces an additional interaction 
parameter $\varepsilon_B'$ characterizing the final-state interactions.

Calculation of the dressed Green's function is analogous to 
the polaron problem considered in Ref.~\cite{Schmidt:2011}, and we obtain the self-energy 
$\Sigma_f^{}$ by convoluting $\mathcal{G}_{0,\uparrow}$ with the many-body $T$-matrix 
$T_{\uparrow f}$. Since the $T$-matrix corresponding to the final-state interactions does not have 
poles away from the real axis~\cite{Schmidt:2011}, we can directly apply spectral 
representation for the $T$-matrix and obtain~\cite{Engelbrecht:1992}
\begin{align}
\label{se_fs}
\im\,\Sigma_{f}^{}(\bq,\om) = \int & \frac{\dk}{(2\pi)^2}  \big[\nb(\om+\eps{\bk}-\mu)+\nf(\eps{\bk}-\mu) 
\big]\,  \notag \\  & \quad \times\, \im\,T_{\uparrow f}^{}(\bq+\bk,\om+\eps{\bk}-\mu).
\end{align}
Using Eq.~\eqref{se_fs}, we calculate the real part using the Kramers-Kronig 
relation~\eqref{kk_rel}. In Ref.~\cite{Schmidt:2011}, the self-energy of the final-state atoms 
was calculated at $T=0$ without invoking the Kramers-Kronig relation. As a check for the 
numerical calculation, we have verified that we reproduce the results of Ref.~\cite{Schmidt:2011} 
in the appropriate limit.


\section{\label{chem_pot}Calculation of the chemical potential}

In this section, we compare two different schemes for calculating the chemical potential of the 
initial-state atoms. The first one is the Nozi\'eres--Schmitt-Rink (NSR) 
approximation~\cite{Nozieres:1985,SchmittRink:1989,Tokumitu:1993,SaDeMelo:1993} and the 
second method is based on the number density given by the dressed Green's 
function~\cite{Serene:1989,Tsuchiya:2009}. The motivation for studying the two 
approximations is the fact that the NSR scheme is numerically much more affordable but expected 
to be accurate only when self-energy corrections are small~\cite{Serene:1989}. We are not aware 
of any explicit comparison of the two schemes for 2D systems. 

Both aforesaid methods start from the thermodynamical potential 
$\Omega[\Sigma,\mathcal{G}]$~\cite{Serene:1989}, where $\Sigma$ and $\mathcal{G}$ are 
self-energy and dressed Green's function, respectively. The density is given by 
$n = - \frac{\partial\Omega}{\partial\mu}$,
and the NSR approximation is obtained by replacing $\Omega[\Sigma,\mathcal{G}] \rightarrow 
\Omega[\Sigma,0]$. A more rigorous approximation suggested by Serene~\cite{Serene:1989} 
corresponds to $\Omega[\Sigma,\mathcal{G}] \rightarrow \Omega[\Sigma,\mathcal{G}_0^{}]$. 
The NSR approximation results in a number equation of the 
form~\cite{SchmittRink:1989,Tokumitu:1993,Serene:1989,Yamada:2000}
\begin{align}
\label{nsr_number_eq}
n = & 2\int\frac{\dq}{(2\pi)^2} \,\nf(\eps{\bq}-\mu) + \notag \\ 
& \quad \quad \quad \frac{\partial}{\partial\mu}\int\frac{\dq}{(2\pi)^2}
\,\mathcal{P}\int_{-\infty}^{+\infty}
\frac{\mathrm{d}\om}{\pi}\,\nb(\om)\,\delta(\bq,\om),
\end{align}
where $\delta(\bq,\om)$ is the phase shift of the $T$-matrix, that is, $\tmat(\bq,\om) = 
\big| \tmat(\bq,\om) \big |\,e^{i\delta(\bq,\om)}$. The total density can be written in terms of the 
Fermi energy as $n=m\varepsilon_F^{}/\pi$. Numerical solution of Eq.~\eqref{nsr_number_eq} 
determines the chemical potential $\mu$ in terms of the Fermi energy $\varepsilon_F^{} = 
k_F^2/2m$. We denote the temperature corresponding to Fermi energy by $T_F^{}$.

A more rigorous alternative to the NSR number equation is a  
full number equation where the density is given by the 
dressed Green's function~\cite{Serene:1989,Perali:2002,Tsuchiya:2009}
\be
\label{numb1}
n = \frac{2}{\beta}\sum_{i\om_n^{}}\int\frac{\dq}{(2\pi)^2}\,\mathcal{G}(\bq,i\om_n^{})\,
e^{i\om_n^{}0^+_{}},
\ee
such that $\mathcal{G}(\bq,i\om_n^{})$ is obtained from the Dyson's equation~\eqref{dyson}. 
Using the spectral representation, we can cast Eq.~\eqref{numb1} 
into 
\be
\label{numb2}
n = 2\int\frac{\dq}{(2\pi)^2} \int_{-\infty}^{\infty}\frac{\mathrm{d}\om}{2\pi}   \,A(\bq,\om)\,\nf(\om).
\ee
The NSR number equation is obtained from Eq.~\eqref{numb1} if $\mathcal{G}(\bq,i\om_n^{})$  
is replaced with~\cite{Serene:1989,Tsuchiya:2009}
\be
\mathcal{G}^{\text{NSR}}_{}(\bq,i\om_n^{}) = \mathcal{G}_0^{}(\bq,i\om_n^{}) + 
\mathcal{G}_0^{}(\bq,i\om_n^{})\Sigma(\bq,i\om_n^{})\mathcal{G}_0^{}(\bq,i\om_n^{}).
\ee
The NSR approximation is therefore formally reliable when corrections due to nonzero 
self-energy are small. 

\vspace{3mm}
\begin{figure}[h!]
\begin{center}
\includegraphics[width=0.4\textwidth]{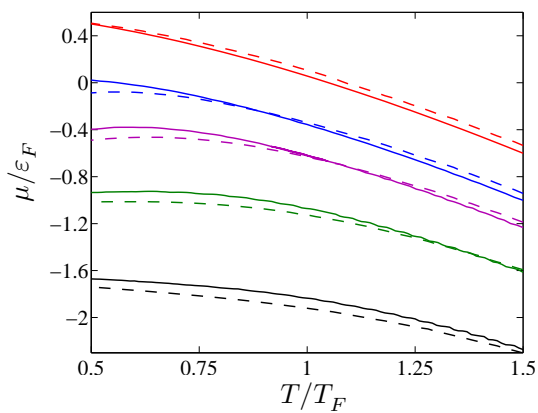}
\end{center}
\caption{\label{mu_fixed_eta} (Color online) Chemical potential as a function of temperature for 
different values of the interaction parameter $\eta$. From top to bottom: $\eta=1.5,\, 0.5,\, 0.25,\, 0,$ 
and $-0.25$. The solid line corresponds to the full number equation [Eq.~\eqref{numb2}] and the 
dashed line is the NSR approximation [Eq.~\eqref{nsr_number_eq}].}
\end{figure}

\begin{figure}[h!]
\begin{center}
\includegraphics[width=0.4\textwidth]{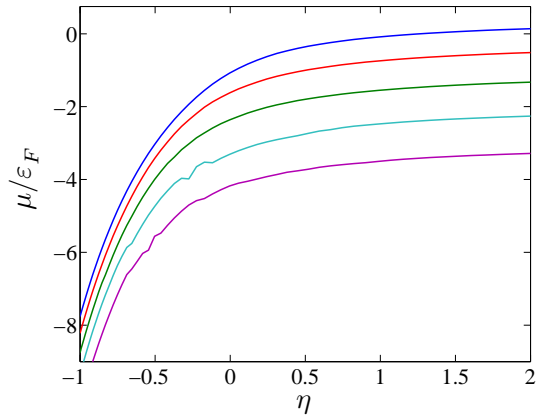}
\end{center}
\caption{\label{mu_fixed_t} (Color online) Chemical potential at different temperatures as a 
function of the interaction parameter $\eta$. From top to bottom: $T/T_F^{}=1.0,\, 1.5,\, 2.0,\, 2.5,$ 
and $3.0$. }
\end{figure}

Before analyzing the 2D case, we note that in 3D (at the critical temperature), the NSR 
approximation and the full number equation yield chemical potentials that are almost 
identical~\cite{Tsuchiya:2009}. In Fig.~\ref{mu_fixed_eta} we compare the results of the NSR 
approximation and the full number equation for a 2D Fermi gas. We present the results in terms 
of a dimensionless interaction parameter~\cite{Feld:2011, Schmidt:2011},
\be
\label{eta_def}
\eta = -\frac{1}{2}\ln(\eb/2\varepsilon_F^{})=\ln(\kf\asc).
\ee 
The 2D unitarity is defined as $\eta=0$ since perturbative expansions in terms of $1/\eta$ 
diverge at this point~\cite{Bloom:1975}. A BCS-type superfluid and a 
Bose-Einstein condensate~(BEC) of tightly bound molecules correspond to $\eta \gg 1$ and 
$\eta \ll -1$, respectively~\cite{Randeria:1989,Bertaina:2011}. We observe that the difference 
between the NSR approximation and the full number equation is surprisingly small and the NSR 
theory is in fact a good approximation for the full number equation in the 
regime of parameters relevant to this work. As one would expect, the NSR 
approximation works best in the weakly interacting regime where $\eta > 1$ (BCS limit). 
We will use, however, the chemical potential computed from the full number 
equation~\eqref{numb2} in the subsequent calculations. In Fig.~\ref{mu_fixed_t}, we show the 
chemical potential as a function of the interaction parameter $\eta$ at different temperatures.

\vspace{3mm}

\section{\label{spec_properties}Spectral functions and quasiparticle spectrum}

We consider two separate cases of strongly interacting Fermi atoms: (A) a balanced 
mixture of spin-$\downarrow$ and spin-$\uparrow$ atoms and (B) an impurity problem where an 
atom in a state $|f\ra$ is immersed in a Fermi sea of spin-$\uparrow$ atoms. In the context of 
rf spectroscopy~\cite{Sommer:2012,Zhang:2012}, case (A) corresponds to the initial state of the 
system and case (B) describes the final-state atoms interacting with spin-$\uparrow$ atoms. 
In case (B) we assume that the impurity only interacts with spin-$\uparrow$ atoms since 
the rf pulse transferring atoms from state $|\downarrow\ra$ to the final-state $|f\ra$ 
simply rotates the atomic spin. 

\subsection{\label{spectral_funcs_initial}Initial balanced mixture}

\begin{figure}[h!]
\begin{center}
\includegraphics[width=0.49\textwidth]{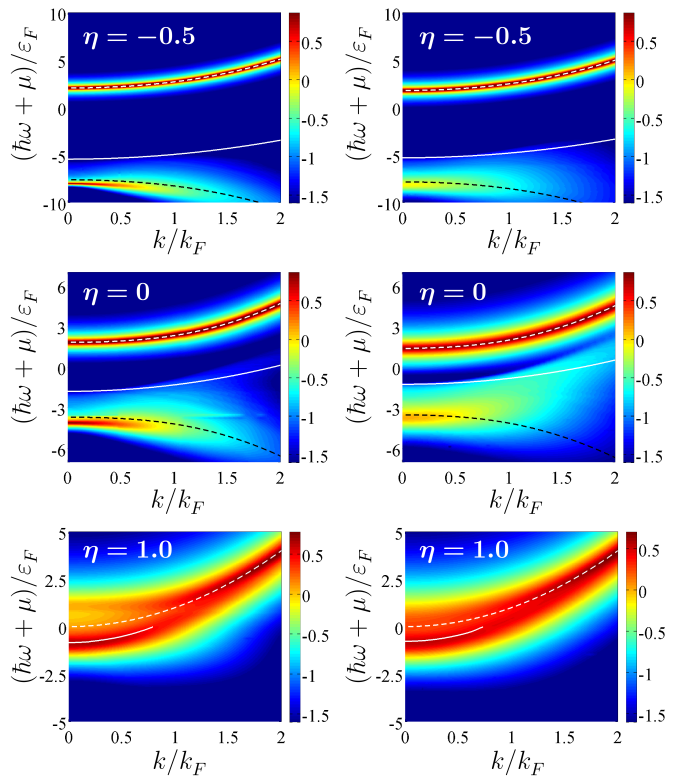}
\end{center}
\caption{\label{spfs} (Color online) Spectral function in the energy-momentum plane. Left panels: 
$T/T_F^{} = 0.5$. Right panels: $T/T_F^{} = 1.0$. The color scale is given by 
$\log_{10}^{} A_{\downarrow}^{}(\bk,\om)$. For $\eta=-0.5$ and $\eta=0$, the dashed lines 
indicate a  BCS-like dispersion given by Eq.~\eqref{ps_gap} and the solid line is the threshold 
given by Eq.~\eqref{threshold1}. For $\eta=1.0$, the dashed line corresponds to 
the free particle dispersion $\eps{\bk} = \bk^2/2m$ and the solid line indicates the dispersion of the 
attractive polaron.}
\end{figure}

The spectral function $A_{\downarrow}^{}(\bk,\om)$ for different values of the interaction parameter 
$\eta$ is shown in Fig.~\ref{spfs} for $T/T_F^{}=0.5$ and $1.0$.  In order to reach 
qualitative  agreement with experimentally measured spectral functions~\cite{Feld:2011}, one 
has to average the spectral function over the inhomogeneous density to account for the free atoms 
residing at the edge of the trap. Such atoms typically give rise to a peak at 
$\omega=-\mu$~\cite{He:2005,Massignan:2008,Chen:2009} (see also Sec.~\ref{ncsu_exp}).
To understand the properties of the resulting rf spectra, we analyze next the different contributions 
to the quasiparticle spectra. For $\eta=-0.5$ and $0$, we find two distinct features in the 
spectral function: a broad incoherent band and a narrow coherent band. This two-band structure 
is a first indication of the pseudogap regime~\cite{Perali:2002, Stewart:2008,Tsuchiya:2009, 
Haussmann:2009,Hu:2010, Gaeblar:2010,Mueller:2011,Chien:2010} since it suggests that 
removing a spin-$\downarrow$ particle from the system corresponds to creating and annihilating 
a pair of quasiparticles as in the usual BCS theory~\cite{Tsuchiya:2009}.

\vspace{3mm}

We observe that the threshold for the incoherent branch in the spectral function 
(for $\eta = -0.5$ and $0$) is approximately given by 
\be
\label{threshold1}
\om_{\text{th}}^{}(\bq) = \epair(\bq) + \mu,
 \ee
where the bound-state energy $\epair(\bq)$ is given by the pole of the many-body $T$-matrix. 
We find that the poles have only a vanishingly small imaginary part. The threshold 
$\om_{\text{th}}^{}(\bq)$ is  obtained from a simple argument: to create a spin-$\downarrow$ 
quasiparticle excitation with momentum $\bq$, one can create a fermion pair with momentum 
$\bq+\bk$ and remove a spin-$\uparrow$ atom with momentum $\bk$. This requires energy 
$E=\epair(\bk+\bq) - (\eps{\bk}-\mua)$ and the threshold is obtained 
for $|\bk|\simeq 0$. In a balanced system,  this results in an energy threshold 
$\om_{\text{th}}^{}(\bq) = \epair(\bq) + \mu$ for the incoherent part of the quasiparticle spectrum. 
We note that a similar argument has been put forward earlier in Ref.~\cite{Mueller:2011}. 

\begin{figure*}[t]
\begin{center}
\includegraphics[width=0.95\textwidth]{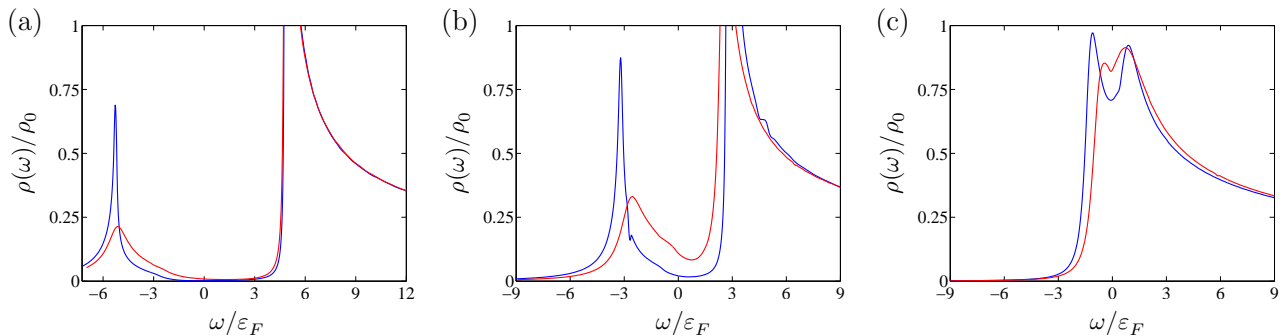}
\end{center}
\caption{\label{dos_fig} (Color online) Density of states  $\rho(\om)$ for (a) $\eta = -0.5$, 
(b) $\eta = 0$, and (c) $\eta = 1.0$. The upper (lower) curves with respect to the left peak 
correspond to $T/T_F^{} = 0.5$ ($T/T_F^{}=1.0$). The density of states is measured in the units 
of density of states for an ideal Fermi gas, $\rho_0^{} = m/\hbar^2$. The parameters are the 
same as in Fig.~\ref{spfs}.}
\end{figure*}

To gain more insight into the two branches of the spectral function for $\eta=-0.5$ and $0$, 
we fit a BCS-like dispersion~\cite{Perali:2002,Haussmann:2009,
Tsuchiya:2009,Gaeblar:2010,Chien:2010}, 
\be
\label{ps_gap}
\varepsilon_{\text{BCS}}^{}(\bk) = \sqrt{(\eps{\bk} + U - \mu)^2 + \Delta_{\text{pg}}^2},
\ee 
to the numerical data. The pseudogap in Eq.~\eqref{ps_gap} is denoted by 
$\Delta_{\text{pg}}^{}$ and $U$ corresponds to a Hartree shift. The spectral function in 
Fig.~\ref{spfs} is shown relative to the chemical potential $\mu$ and therefore the two bands 
are described by an effective dispersion,  
$E_{\pm}^{}(\bk) = \mu \pm \varepsilon_{\text{BCS}}^{}(\bk)$~\cite{Haussmann:2009,Chen:2009}. 
The spectral function itself is peaked around the effective BCS-like dispersion given by 
$\varepsilon_{\pm}^{}(\bk) = \pm \varepsilon_{\text{BCS}}^{}(\bk)$. Results from a least squares fit 
are shown in Table~\ref{bcs_fit} and we observe that the pseudogap $\Delta_{\text{pg}}^{}$ 
decreases with increasing temperature and $\eta$. In general, we find that the coherent 
particle branch tends to be more accurately described by the BCS-like dispersion than the hole 
branch. From Fig.~\ref{spfs} and  Table~\ref{bcs_fit}, we conclude that for $\eta=-0.5$ and 
$\eta=0$, the pseudogap regime extends at least up to $T=T_F^{}$ if we define it to correspond 
to a regime where $\Delta_{\text{pg}^{}}$ is nonzero. We note that the true pairing gap 
corresponding to the superfluid phase of a 2D Fermi gas has been calculated at $T=0$ in 
Ref.~\cite{Bertaina:2011}. 

\vspace{3mm}
\begin{table}[h!]
\begin{tabular}{l | c | c | r}
$\eta$ & $T/T_F^{}$ & $\Delta_{\text{pg}^{}}/\varepsilon_F^{}$ & $U/\varepsilon_F^{}$ \\[3pt]
\hline \hline
-0.5 \hspace{4pt } & \hspace{5pt } 0.5 \hspace{5pt } & \hspace{6pt } 3.90 \hspace{6pt } & 
0.08 \\[3pt]
-0.5 & 1.0 & 3.64 & 0.11 \\[3pt]
0 & 0.5 & 2.57 & 0.07 \\[3pt]
0 & 1.0 & 2.15 & 0.10
\end{tabular}
\caption{\label{bcs_fit} Pseudogap $\Delta_{\text{pg}^{}}$ and Hartree shift $U$ arising from a 
least-squares fit of a BCS-like dispersion to the numerical data in Fig.~\ref{spfs}.}
\end{table}

In the context of 3D Fermi gases, the pseudogap regime has been investigated by analyzing the 
density of states~\cite{Tsuchiya:2009,Mueller:2011}. For 2D systems, the corresponding DOS is 
given by 
\be
\rho(\om) = \int\frac{\dbk}{(2\pi)^2}\,A(\bk,\om). 
\ee
In Fig.~\ref{dos_fig}, we show the DOS corresponding to Fig.~\ref{spfs}. The density of states is 
measured with respect to the corresponding density of states of a noninteracting Fermi gas. In 
two spatial dimensions, ideal-gas DOS is a constant, $\rho_0^{} = m/\hbar^2$. In terms of DOS, the 
pseudogap regime can be identified as a regime where DOS becomes strongly suppressed at 
zero energy (with respect to chemical potential)~\cite{Tsuchiya:2009,Mueller:2011}. Using  
this criterion, we observe that the system is in the pseudogap regime 
for $\eta = -0.5$ and $\eta=0$ at least up to temperature $T=T_F^{}$; see Fig.~\ref{dos_fig}. 
On the other hand,  the pseudogap practically vanishes for $\eta=1.0$ at  $T/T_F^{} = 0.5$ and 
$1.0$.  Another characteristic of the pseudogap regime is ``backbending" of the 
quasiparticle dispersion in the single-particle spectral function near the Fermi wave 
vector~$\kf$~\cite{Perali:2011}. From Fig.~\ref{spfs}, we observe that the backbending near 
$\kf$ is clearly manifested in the spectral function for $\eta = -0.5$ and $\eta=0$ at $T/T_F^{} = 0.5$. We note that the backbending in the normal state of 
a Fermi gas has been discussed in Ref.~\cite{Schneider:2010}, but in this case the backbending 
is expected to take place at $|\bk| \gg \kf$. The BCS-like dispersion, suppression of DOS near zero 
energy, and backbending of the dispersion suggest that the gas is in the pseudogap regime 
for $\eta = -0.5$ and $\eta=0$  at least up to $T=T_F^{}$.

For $\eta = 1.0$, we find that the $T$-matrix does not have any poles (for $T/T_F^{}=0.5$ and 
$1.0$) as the bound state is pushed against the two-particle continuum in the 
molecule spectral function. From Fig.~\ref{spfs}, we observe that the spectral function (for 
$\eta=1.0$) at large momenta corresponds to free-particle-like excitations. The low-momentum 
part is more interesting since it is best described in terms of non-interacting polarons. In this 
regime ($|\bk| < \kf$), the spectral weight is centered around energies that coincide with the 
dispersion of the attractive polaron. We define the (attractive) polaron energy as in 
Refs.~\cite{Schmidt:2011, Combescot:2007,Massignan:2011}, that is, we consider a single 
spin-$\downarrow$ impurity embedded to a Fermi sea of spin-$\uparrow$ atoms. The polaron 
dispersion follows from equations
\begin{align}
\label{pol1}
&\om(\bk)  + \mu_{\downarrow}' - \eps{\bk} - \re\,\Sigma_{\downarrow}^{}\big(\bk,\om(\bk)\big) = 0, \\
\label{polaron_dispersion}
&\om_{p}^{}(\bk) = \om(\bk) + \mu_{\downarrow}',
\end{align}
where $\mu_{\downarrow}'$ is tuned such that state $|\downarrow\ra$ is empty. To ensure that 
the calculation is self-consistent, we set $\mud = \mu$, where $\mu$ is solved from the number 
equation~\eqref{numb2} corresponding to the original balanced system. Equations~\eqref{pol1} 
and~\eqref{polaron_dispersion} generalize the zero-temperature analysis~\cite{Schmidt:2011} 
to finite temperatures. Moreover, at $T=0$, Eqs.~\eqref{pol1} and~\eqref{polaron_dispersion} are 
equivalent to the analysis based on a variational wave function for the 
polaron~\cite{Zollner:2011,Parish:2011}.

Our analysis suggests that outside the pseudogap regime,  
even a balanced 2D Fermi gas can be effectively described as a gas of noninteracting polarons.  
For 2D systems, the polaron description has been used by Zhang {\it et al.}~\cite{Zhang:2012} as 
a possible explanation for their experimental observations. For 3D systems, similar speculations 
have appeared earlier in Ref.~\cite{Schirotzek:2009}. Our findings thus support the scenario put 
forward in Ref.~\cite{Zhang:2012} and indicate a crossover (at fixed temperature) from a polaron 
gas for small attraction to a pseudogap regime at strong attraction.

\subsection{\label{fs_spectral}Final state -- impurity}

The quasiparticle excitations of the final state have been thoroughly discussed in 
Ref.~\cite{Schmidt:2011} at $T=0$. The main features in the final-state spectral function 
correspond to the attractive and the repulsive polaron~\cite{Massignan:2011, 
Schmidt:2011, Schmidt:2011b}, while the contribution from a bound state carries only a small 
spectral weight. Here we briefly discuss the qualitative changes that the 
finite temperature induces. As in Ref.~\cite{Schmidt:2011}, we first assume that the initial mixture 
of spin-$\uparrow$ and spin-$\downarrow$ atoms is noninteracting. The chemical potential  
is then given by the ideal gas expression 
$\mu_0^{}(T) = \kb T\,\ln\big(e^{\varepsilon_F^{}/\kb T} - 1\big)$. 
The finite temperature results in two qualitative changes: the attractive polaron acquires a finite 
lifetime and the molecule-hole continuum merges with the attractive polaron branch. Furthermore, 
as the temperature increases, the spectral weight is distributed more equally between attractive 
and repulsive polarons. We note that the properties of 2D polarons have recently been studied 
experimentally in Refs.~\cite{Frohlich:2011, Koschorreck:2012} and the experimental data 
agrees reasonably well with the non-self-consistent $T$-matrix 
calculation~\cite{Schmidt:2011,Koschorreck:2012}. 

\vspace{3mm}
\begin{figure}[h!]
\begin{center}
\includegraphics[width=0.49\textwidth]{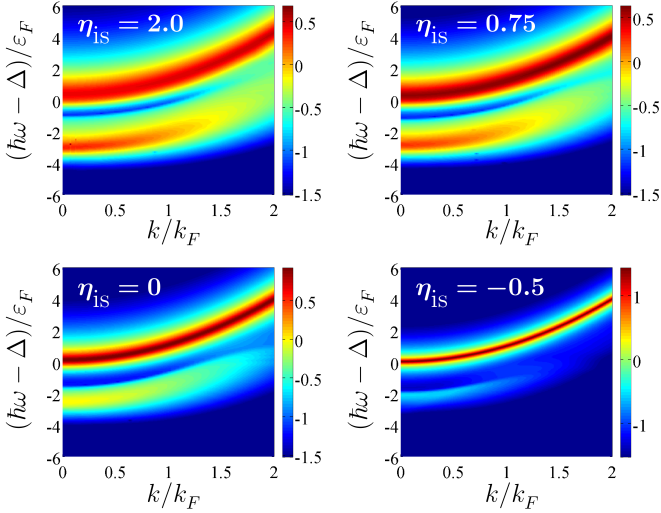}
\end{center}
\caption{\label{polaron_ft} (Color online) Spectral function of the final-state atom (impurity) when 
the majority atoms are dressed by strong initial-state interactions. The color scale corresponds to 
 $\log_{10}^{} A_{f}^{}(\bk,\om)$. The strength of 
the initial-state interaction is denoted by $\eta_{\text{is}}^{}$ in the figure. The temperature is 
set to $T=T_F^{}$ and interactions between the final-state atom and spin-$\uparrow$ atoms 
correspond to $\eta_{\text{fs}}^{}=0$. The level splitting between states $|\uparrow\ra$ and 
$|f\ra$ is denoted by $\Delta$ and is assumed to be positive.}
\end{figure}

Next we consider the case where the impurity (final-state atom) is dynamically created 
from an interacting, balanced gas of spin-$\uparrow$ and spin-$\downarrow$ atoms. The 
chemical potential is now determined by the interacting initial state and it is solved from the 
number equation~\eqref{numb2}. In Fig.~\ref{polaron_ft}, we show the spectral function of the final 
state atoms for different initial-state interactions. For illustrative purposes, we fix the final-state 
interaction to $\eta_{\text{fs}}^{} = 0$. In contrast to the experiments in 
Refs.~\cite{Sommer:2012,Zhang:2012}, this final-state interaction is in the strongly interacting 
regime. With increasing initial-state attraction, the initial state becomes increasingly paired,   
and the added impurity is less likely to distort the cloud of majority atoms. Thus the impurity 
behaves more and more like a free particle as one approaches the BEC limit. 
In Fig.~\ref{polaron_ft}, this is manifested as a suppression of the spectral weight associated with 
the polaron states. 

\section{\label{rf}Radio-frequency spectroscopy}

We consider radio-frequency spectroscopy in which a balanced mixture of 
atoms in hyperfine states $| i_1^{}\ra \equiv |\uparrow\ra$ \\ and $| i_2^{}\ra 
\equiv |\downarrow\ra$ is coupled to rf photons inducing a transition $|\downarrow\ra 
\rightarrow |f\ra$~\cite{Chen:2009b}. Within the linear response framework, one obtains  
\be
\label{irf}
\irf(\oml) = -\pi\omrff\,\im\,\chi_{R}^{}(\bk=0,-\oml-\mua+\mu_f^{}).
\ee
At finite temperatures, the retarded correlation function $\chi_{R}^{}$ can be computed from the 
corresponding time-ordered correlation function as~\cite{Doniach:1974}
\be
\label{ret_vs_timeord}
\im\,\chi_{R}^{}(\bk,\om) = \tanh(\frac{1}{2}\beta\om)\, \im\,\chi(\bk,\om),
\ee
where
\begin{align}
&\chi(\br-\br',t-t') =  \notag \\
& -i\theta(t-t')\la[\psi_f^{\da}(\br,t)\psi_\downarrow^{}(\br,t),\psi_\downarrow^{\da}(\br',t')
\psi_f^{}(\br',t')]\ra.
\end{align}
Since we evaluate the correlation function for $\om = -\oml-\mu+\mu_f^{} \ll -\varepsilon_F^{}$, 
we have $\tanh(\frac{1}{2}\beta\om)\approx-1$.  We neglect the vertex 
corrections~\cite{Perali:2008, Pieri:2009,Pieri:2011} and obtain
\begin{align}
\label{ph_se}
\chi(\bk = & 0,  i\Omega_n) = \notag \\
&\frac{1}{\beta}\sum_{i\omega_m^{}}\int\frac{\dbq}{(2\pi)^2}
\mathcal{G}_{f}^{}(\bq,i\omega_m^{})\mathcal{G}_{\downarrow}^{}(\bq,i\Omega_n^{}+
i\omega_m^{}).
\end{align}

Using the spectral representation for the Green's functions, the analytic continuation can be 
performed exactly and we obtain
\begin{align}
\label{rf_current}
\irf(\om) = &\frac{\pi}{2}|\omrf|^2\int\frac{\dbq}{(2\pi)^2}\int\frac{\dz}{2\pi}\,A_{\downarrow}^{}(\bq,
z-\om-\Delta_{\uparrow f}^{}-\mua) \notag \\
&\times A_f^{}(\bq,z)\big[ \nf(z-\om-\Delta_{\uparrow f}^{}-\mua)-\nf(z)\big],
\end{align}
where $\Delta_{\uparrow f}^{}$ is the level splitting between states $|\downarrow\ra$ and 
$|f\ra$ and we have defined the detuning $\om = \om_{\mathrm{rf}}^{} - \Delta_{\uparrow f}^{}$. 
If interactions between states $|\uparrow\ra$ and $|f\ra$ are negligible, then the final-state spectral 
function becomes $A_f^{}(\bq,z) = 2\pi\delta(z-\eps{\bq}-\Delta_{\uparrow f}^{})$. Here we have 
set $\mu_f^{}=0$ since the final state is initially empty. We have also taken explicitly into account 
the level splitting $\Delta_{\uparrow f}^{}$. The rf current takes the form
\be
\label{no_fs_int}
\irf(\om) = \frac{\pi|\omrf|^2}{2}\int\frac{\dbq}{(2\pi)^2}\,A_{\downarrow}^{}(\bq,\eps{\bq}-
\mu_{\downarrow}^{}-\om)
\,\nf(\eps{\bq}-\mu_{\downarrow}^{}-\om),
\ee
where we have neglected the term $\nf(\eps{\bq}+\Delta_{\uparrow f}^{})$ corresponding to the 
occupation of the final state. At finite temperatures this term is, in principle, nonzero but we 
assume $\Delta_{\uparrow f}^{}  \gg \varepsilon_F^{}$, which renders 
$\nf(\eps{\bq}+\Delta_{\uparrow f}^{})$ negligible. In  
the current experiments~\cite{Zhang:2012,Sommer:2012}, the Zeeman splitting between the 
relevant hyperfine spin states is $\Delta_{\uparrow f}^{} \sim 50$ MHz~\cite{Bartenstein:2005}, 
whereas the Fermi energy is in the range 
$\varepsilon_F^{} \sim 10-250$~kHz~\cite{Zhang:2012,Sommer:2012}. 
Hence, we expect that neglecting the occupation of the final state is justified even at temperatures 
$T\sim T_F^{}$. 

We note that the more general form of the rf current in Eq.~\eqref{rf_current} is 
phenomenological since all vertex corrections are excluded. Therefore our formalism does not 
include bound-to-bound transitions~\cite{Schunck:2008,Pieri:2009}.
Since Eq.~\eqref{rf_current} contains information about the final-state polarons, we will use it to 
analyze the NCSU experiment. 

\subsection{Quasi-2D geometry and experimental parameters}

To compare our results to the experiments~\cite{Sommer:2012,Zhang:2012}, we 
need to establish a connection between the dimensionless interaction parameter $\eta$ and an  
external magnetic field $B$ which is used to tune the interactions via the Feshbach 
resonance. Since the experiments are performed in a quasi-2D geometry, we calculate 
the two-body binding energy $\eb$ in terms of the 3D scattering length 
$\ass$ using the quasi-2D form of the $T$-matrix~\cite{Petrov:2001,Bloch:2008,Pietila:2012},
\be
\mathcal{T}_0(\om) = \frac{2\sqrt{2\pi}/m}{\ell_z^{}/\ass - \mathcal{F}(-\om/\om_z^{})},
\ee
where function $\mathcal{F}(z)$ is given by 
\be
\mathcal{F}(z) = \int_0^{\infty}
\,\frac{\mathrm{d}u}{\sqrt{4\pi}}\,\frac{1}{u^{3/2}}\bigg( 1 - 
\frac{e^{-zu}}{\sqrt{\frac{1}{2u}(1-e^{-2u})}}\bigg).
\ee
The trap frequency along the tightly confined direction is denoted by $\om_z^{}$ and the length 
scale associated with the confinement is given by $\ell_z^{} = \sqrt{1/m\om_z^{}}$ for atoms with 
mass $m$. The two-body binding energy $\eb$ corresponds to a pole of the vacuum $T$-matrix 
and is determined by the equation (we assume $\eb>0$)
\be
\label{2d_eb}
\ell_z^{}/\ass - \mathcal{F}(\eb/\om_z^{}) = 0.
\ee
We solve Eq.~\eqref{2d_eb} numerically to obtain $\eb = \eb(\ell_z^{}/\ass)$. To convert the 
external magnetic field to the 3D scattering length $\ass$, we use the parameters of 
Ref.~\cite{Bartenstein:2005}. The two-body binding energy is converted to the dimensionless 
interaction parameter $\eta$ using Eq.~\eqref{eta_def}.

\subsection{\label{mit_exp}MIT experiment~\cite{Sommer:2012}}

The MIT experiment~\cite{Sommer:2012} is performed using the three lowest sublevels 
$|1\ra$, $|2\ra$, and $|3\ra$ on the $^6$Li ground-state manifold~\cite{Bartenstein:2005}.
The set of initial states is given by $\{|\uparrow\ra, |\downarrow\ra\}  = \{|1\ra,|3\ra\}$, and the final 
state corresponds to $|f\ra = |2\ra$. Interactions between atoms in the two initial states are 
enhanced by a Feshbach resonance at $B=690.4$~G. The initial state $|\downarrow\ra$ 
coupled with the final state $|f\ra$ is chosen such that the final-state interactions are 
minimized~\cite{Sommer:2012}. In the MIT experiment, the two-body binding energy of the final 
state dimer is much larger than the initial-state dimer energy, $\varepsilon_B' \gg \eb$. As in the 
3D case~\cite{Schunck:2008}, the final-state interactions are important only if 
$\eta_{\uparrow f}^{}$ corresponding to $\varepsilon_B'$ is close to the 2D unitarity $\eta=0$. 
The MIT experiment is outside this regime and hence we neglect the final-state interactions 
altogether. We note that the final-state interactions are important for the bound-to-bound 
transition in the rf spectrum~\cite{Langmack:2012}, but we do not consider this transition in the 
present work. Although the MIT experiment studied the whole dimensional crossover from 3D to 
2D, we concentrate on the 2D limit of the experiment. Using the methods of 
Ref.~\cite{Pietila:2012}, it is possible to study rf spectroscopy across the whole dimensional 
crossover. Due to the multiple bands involved with this formalism, the calculation becomes 
numerically more demanding and we postpone such studies for future research.

\begin{table}[h!]
\begin{tabular}{c | c }
B [G] & $\eta_{\uparrow\downarrow}^{}$  \\[3pt]
\hline \hline
690.7 \hspace{2mm} & \hspace{2mm}  -0.504   \\[3pt]
720.7 \hspace{2mm} & \hspace{2mm} -0.364 
\end{tabular}
\caption{\label{mit_eta} Interaction parameter $\eta$ corresponding to the 2D limit of the 
MIT experiment. The values of the experimental parameters can be found from 
Ref.~\cite{Sommer:2012}. To convert the external magnetic field to the 3D scattering length 
$\ass$, we have used the parameters from Ref.~\cite{Bartenstein:2005}.}
\end{table}

The dimensionless interaction parameter $\eta$ for the initial-state interactions is shown in 
Table~\ref{mit_eta} for the relevant values of the external 
magnetic field. The temperature at the MIT experiment was estimated to be of the order of 
$T_F^{}$ and the peak Fermi energy was approximately 
$\varepsilon_F^{}=h\times10$ kHz~\cite{Sommer:2012}. 
We note that the data corresponding to the deepest optical lattice (the regime of our interest) 
was deeply in the 2D regime since $\varepsilon_F^{}/\hbar\om_z^{} \approx 0.04$. Thus   
we expect that this limit is well described by our 2D calculation.

\begin{figure}[h!]
\begin{center}
\includegraphics[width=0.48\textwidth]{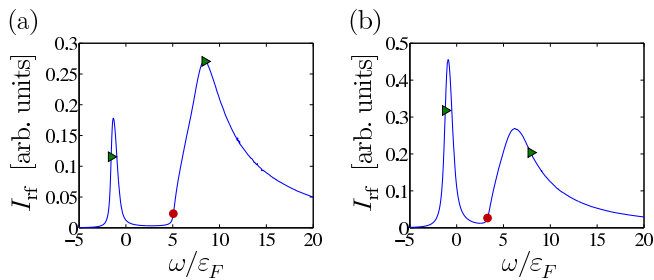}
\end{center}
\caption{\label{rf_mit_comp} (Color online) Radio-frequency spectra  
corresponding to the MIT experiment~\cite{Sommer:2012}. The external magnetic field is (a) 
$B = 690.7$~G and (b) $B=720.7$~G. The dimensionless interaction parameter $\eta$ for the 
initial-state interactions is given in Table~\ref{mit_eta}. The temperature is (a) $T/T_F^{} = 1.5$ 
and (b) $T/T_F^{} = 2.0$. The dots indicate the threshold given by Eq.~\eqref{threshold} and the 
triangles arise from the BCS-like dispersion at zero momentum; see Eqs.~\eqref{ps_gap} 
and~\eqref{peak_location}. }
\end{figure}

Since the final-state interactions mainly affect the tails of the rf spectra and round off 
the onset of the pairing peak~\cite{Sommer:2012,Langmack:2012}, the qualitative features 
of the rf spectra measured in Ref.~\cite{Sommer:2012} can be understood by studying the spectral 
function of the initial state. In the absence of final-state interactions, Eq.~\eqref{no_fs_int} suggests  
that the broad peak (``pairing peak") at positive detuning arises from the incoherent part of the 
spectral function corresponding to the dissociation of the preformed pairs (see Fig.~\ref{spfs}). 
In Sec.~\ref{spectral_funcs_initial} we showed that the threshold for the incoherent part 
is approximately given by $\om_{\text{th}}^{}(\bq) = \epair(\bq) + \mu$. 
Using Eq.~\eqref{no_fs_int} we obtain a threshold 
\be
\label{threshold}
\om_0^{} = -\om_{\text{th}}^{}(\bq=0) - \mu = -\epair(\bq=0) - 2\mu.
\ee 
Thus the onset of the pairing peak is related to the binding energy of the paired fermions. From 
Fig.~\ref{rf_mit_comp}, we observe that the onset of the incoherent peak is indeed given quite 
accurately by $\om_0^{}$. 

In the pseudogap regime, the spectral function is peaked around an effective BCS-like 
dispersion and Eq.~\eqref{no_fs_int} can be used to find the corresponding frequencies in the 
rf spectrum,
\be
\label{peak_location}
\om_{\pm}^{}(\bq) = \eps{\bq} \pm \varepsilon_{\text{BCS}}^{}(\bq) - \mu.
\ee
However, evaluating $\om_{\pm}^{}(\bq)$ at $\bq=0$ does not directly give the peak location 
since the spectral function is integrated over momentum; see Fig.~\ref{rf_mit_comp}. 
Furthermore, we note that the quasiparticle dispersion starts to deviate from the BCS form for 
$T/T_F^{} = 2.0$ in Fig.~\ref{rf_mit_comp}(b).

The experimentally measured binding energies (the onset of the pairing peak) 
were found to agree reasonably well with the energies of the two-body bound states in 
quasi-2D systems~\cite{Sommer:2012,Orso:2005}. Our calculation for the chemical 
potential at the regime of the MIT experiment ($T\gtrsim T_F^{}$ 
and $\eta< 0$) predicts $\mu$ to be large and negative. Hence, the polarization operator 
$\chi_0^{\text{reg}}(\bq,\om)$ in Eq.~\eqref{chi_reg} gives a negligible contribution to the many-body 
$T$-matrix. In this limit, the poles of the $T$-matrix can be solved exactly and we obtain 
$\epair(\bq) = -\eb - 2\mu + \half\eps{\bq}$. The onset of the pairing peak 
becomes $\om_0^{}(\bq) = \eb - \half\eps{\bq}$, which explains why the experiment is in  
good agreement with the two-body calculation. In order to observe many-body effects, one 
should either consider lower temperatures or quench the system rapidly to the strongly interacting 
regime as suggested in Ref.~\cite{Pietila:2012}. 

\begin{figure}[h!]
\begin{center}
\includegraphics[width=0.48\textwidth]{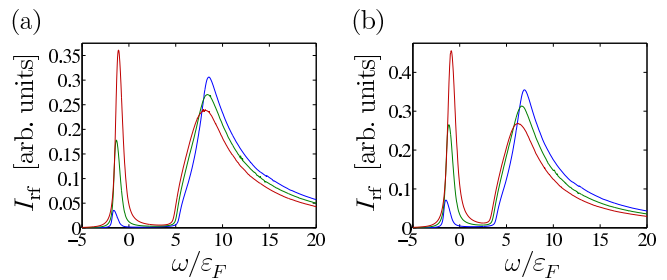}
\end{center}
\caption{\label{rf_mit_temp} (Color online) Radio-frequency spectrum at different temperatures 
corresponding to the 2D limit of Ref.~\cite{Sommer:2012}. The external magnetic field is (a) 
$B = 690.7$~G and  (b) $B=720.7$~G. The dimensionless interaction parameter $\eta$ for the 
initial-state interactions is shown in Table~\ref{mit_eta}. The temperatures are given by 
$T/T_F^{} = 1.0, 1.5,$ and $2.0$ (from top to bottom with respect to the peak on the right-hand 
side) in both panels.}
\end{figure}

To probe the temperature dependence of the rf spectra, we show the calculated rf current at 
different temperatures in Fig.~\ref{rf_mit_temp}. We find the best qualitative agreement with 
Ref.~\cite{Sommer:2012} for temperatures $T/T_F^{} \simeq 1.5$ at the Feshbach resonance 
($B=690.7$ G) and $T/T_F \simeq 2.0$ on the BCS side of the resonance ($B=720.7$ G). To 
reach a quantitative agreement with the experimental data, the rf signal needs to be averaged 
over the inhomogeneous density. Furthermore,  one has to use the temperature and the Fermi 
energy as fitting parameters since they were not exactly known in Ref.~\cite{Sommer:2012}.  

Since the spectral weight carried by the coherent branch of the spectral function increases with 
increasing temperature (see Fig.~\ref{spfs}), the corresponding peak in the rf spectrum at 
negative detuning becomes stronger with increasing temperature (Fig.~\ref{rf_mit_temp}). 
Within our theory, this peak arises from thermally excited quasiparticles and corresponds to 
unpaired atoms. For the fairly high temperatures required to obtain a qualitative 
agreement with the MIT data, the depletion in the density of states is qualitatively the same as in 
Fig.~\ref{dos_fig}(a). The gas is therefore at the crossover region between the pseudogap regime 
and the regime of a normal gas of bosonic molecules~\cite{Watanabe:2010}.

\subsection{\label{ncsu_exp}NCSU experiment~\cite{Zhang:2012}}

In contrast to the MIT experiment, the NCSU experiment found resonances in the rf spectrum 
that do not correspond to paired fermions but transitions between polaronic states. Our analysis 
in Sec.~\ref{spectral_funcs_initial} suggests that a balanced gas (the initial state) in the weakly 
attractive regime can be considered as a gas of noninteracting polarons. Since the atoms in the 
final state can be described in terms of polarons (see Sec.~\ref{fs_spectral}), the 
experimental picture of transitions between polaronic states arises naturally.

\vspace{3mm}
\begin{table}[h!]
\begin{tabular}{l | c | r}
B [G] & $\eta_{\uparrow \downarrow}^{}$ & $\eta_{\downarrow f}^{}$ \\[3pt]
\hline \hline
719 \hspace{4pt }& \hspace{4pt }-0.330 \hspace{4pt } & \hspace{4pt } 1.626 \\[3pt]
810 & 0.967 & 1.767 
\end{tabular}
\caption{\label{ncsu_params} Dimensionless interaction parameters corresponding to the NCSU 
experiment~\cite{Zhang:2012}. initial-state interactions are denoted by 
$\eta_{\uparrow \downarrow}^{}$, and $\eta_{\downarrow f}^{}$ corresponds to the final-state 
interactions. For other magnetic fields and trap depths considered in 
Ref.~\cite{Zhang:2012}, the initial- and the final-state interactions are similar and give  
qualitatively the same results.}
\end{table}

We first analyze the line shape of the rf spectrum in a homogeneous system for the parameters of 
the NCSU experiment. The dimensionless parameters $\eta_{\uparrow \downarrow}^{}$ and 
$\eta_{\uparrow f}^{}$ characterizing the initial- and final-state interactions are shown in 
Table~\ref{ncsu_params}. We use the values reported in Ref.~\cite{Zhang:2012} for the initial 
and the final-state dimer  energies and determine the interaction parameter $\eta$ using 
Eq.~\eqref{eta_def}. Since the final-state interactions are weak, we expect that our 
phenomenological model in Eq.~\eqref{rf_current} is a good description of the dressed final-state 
atoms.

\begin{figure}[h!]
\begin{center}
\includegraphics[width=0.48\textwidth]{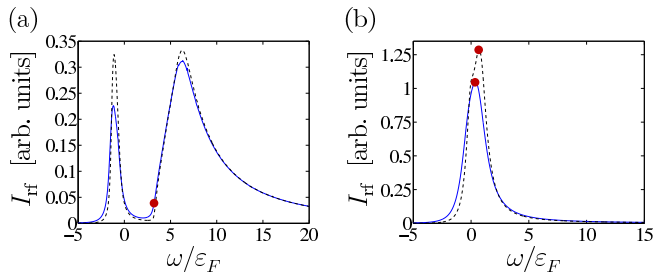}
\end{center}
\caption{\label{rf_ncsu} (Color online) Radio-frequency spectrum corresponding to the NCSU  
experiment~\cite{Zhang:2012} for (a) $B = 719$~G and  
(b) $B=810$~G (solid lines). The dimensionless interaction parameters for the initial- and 
the final-state interactions are shown in Table~\ref{ncsu_params}. The temperature is 
$T/T_F^{} = 1.5$ in both panels. The dashed line denotes the rf spectrum in the absence of final 
state interactions. In panel (a), the dot indicates the threshold given by Eq.~\eqref{threshold},  
and in panel (b), the dots correspond to the polaron threshold; see the main text. The single peak 
structure in panel (b) persists down to $T/T_F^{} = 0.5$, which is the lowest temperature 
considered in this work.}
\end{figure}

The rf spectra corresponding to the parameters of Table~\ref{ncsu_params} are shown in 
Fig.~\ref{rf_ncsu}. For strong attraction [Fig.~\ref{rf_ncsu}(a)], the pairing peak at positive 
detuning $\om$ is again characterized by the threshold frequency $\om_0^{}$ given by 
Eq.~\eqref{threshold}. The sharp peak around zero detuning arises from the unpaired 
atoms as in the MIT experiment. We find that the inclusion of the final-state interactions tends to 
lower the peak corresponding to the unpaired atoms. A crucial difference between 
Figs.~\ref{rf_ncsu}(a) and~(b) is the absence of a pole in the $T$-matrix for 
the parameters of Fig.~\ref{rf_ncsu}(b). Since most of the data presented in 
Ref.~\cite{Zhang:2012} fall to the same regime of initial- and final-state interactions as 
Fig.~\ref{rf_ncsu}~(b), the measured rf spectra cannot be explained in terms of 
confinement-induced molecules. 

Based on the analysis in Sec.~\ref{spectral_funcs_initial}, we argue that the origin of the single 
peak in Fig.~\ref{rf_ncsu}(b) is related to the effective description of the gas in terms of 
noninteracting polarons. In the absence of final 
state interactions, we expect the location of the peak to be given by 
$\om_0^{} = -\om_p^{\text{(is)}}(\bk=0)$, where $\om_p^{\text{(is)}}(\bk)$ is the initial-state 
polaron energy given by Eq.~\eqref{polaron_dispersion}. Furthermore, when the final-state 
interactions are included, the free-particle dispersion of the final-state atoms is effectively 
replaced by the final-state polaron dispersion, and we obtain an estimate for the peak location:  
$\om_0' = -\om_p^{\text{(is)}}(\bk=0) + \om_p^{\text{(fs)}}(\bk=0)$, where 
$\om_p^{\text{(fs)}}(\bk=0)$ is the energy of the final-state polaron.
From Fig.~\ref{rf_ncsu}~(b), we observe that the estimates $\om_0^{}$ and $\om_0'$ are indeed 
quite accurate. 

Similarly to the MIT experiment~\cite{Sommer:2012}, the NCSU experiment has a dual peak 
structure in the measured rf spectra~\cite{Zhang:2012}. Since our calculation for a 
homogeneous system predicts only a single peak, we argue that the second peak is due to 
the averaging of the rf signal over an inhomogeneous density. As in 3D systems~\cite{He:2005, 
Massignan:2008,Chen:2009}, the auxiliary peak arises from unpaired atoms in the low-density 
region at the edge of the trap. To show this explicitly, we average the 
rf signal over the inhomogeneous density using the local density approximation 
(LDA)~\cite{Chiofalo:2002}. Within the 
LDA, the local chemical potential is given by 
$\mu(\br) = \mu- \frac{1}{2}m\omega_{\perp}^2 r_{}^2$. The local density $n_{\text{LDA}}^{}(\br)$ 
is calculated using the number equation~\eqref{numb2} and the density averaged rf signal is 
obtained as $I_{\text{rf}}^{\text{LDA}}(\om) =  \int \text{d}\br\,n_{\text{LDA}}^{}(\br) 
I_{\text{rf}}^{}(\om,\br)$, where $I_{\text{rf}}^{}(\om,\br)$ is the rf signal given by 
Eq.~\eqref{rf_current} for the local chemical potential $\mu(\br)$.

\begin{figure}[h!]
\begin{center}
\includegraphics[width=0.48\textwidth]{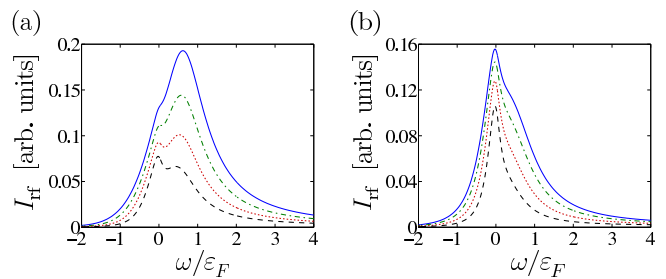}
\end{center}
\caption{\label{ncsu_trap} (Color online) Trap-averaged rf signal corresponding to (a)  
$T/T_F^{} = 0.5$ and (b) $T/T_F = 1.0$. The dimensionless interaction parameters are 
$\eta_{\uparrow\downarrow}^{} = 0.967$ and $\eta_{\downarrow f}^{} = 1.767$. The rf current is scaled by the total number of particles which is given by (a) $N_{\text{tot}}^{} = 2234, 
1791, 1476, 1214$, and (b) $N_{\text{tot}}^{} = 2073, 1810, 1494, 1190$  (from top to bottom). }
\end{figure}

Rather than attempting to exactly reproduce the experimentally measured lineshapes, we use the 
LDA to confirm that the auxiliary peak in the rf spectrum near the zero detuning is due to trap 
average. To this end, we fix the initial- and the final-state interaction strengths to the values given 
in Table~\ref{ncsu_params} and take the Fermi energy $\varepsilon_F^{} = \om_{\perp}^{} 
\sqrt{N_{\text{tot}}^{}}$ to be $114.5$~kHz; see Ref.~\cite{Zhang:2012}. The total number of 
atoms is given by $N_{\text{tot}}^{} = \int\text{d}\br \, n_{\text{LDA}}^{}(\br)$ and the peak value of 
the chemical potential can be tuned to reach a desired number of 
atoms~\cite{Massignan:2008,Chiofalo:2002}. In Fig.~\ref{ncsu_trap}, we show the rf current for 
different numbers of atoms. We implicitly assume that the trap frequency is tuned such 
that $\varepsilon_F^{}$ remains constant. 

\vspace{3mm}

On a qualitative level, our theoretical calculation agrees with the picture put forward in 
Ref.~\cite{Zhang:2012}: in the regime of the experimental parameters, the rf spectrum is best 
described in terms of transitions between noninteracting polaronic states. In general, we find that 
the location of the right-hand side peak in the trap-averaged rf signal [Fig.~\ref{ncsu_trap}~(a)] is 
not directly related to the polaron energy corresponding to the peak 
chemical potential. Therefore, locally resolved rf spectroscopy~\cite{Schirotzek:2008} is desirable 
to confirm the polaron picture. Within the appropriate temperature range, our LDA calculation  
qualitatively reproduces the double-peak structure observed in the experiment; see 
Fig.~\ref{ncsu_trap}(a). The auxiliary peak due to the trap average becomes more pronounced at 
high temperatures and renders the polaron peak indistinguishable if the total number of atoms 
is too small [Fig.~\ref{ncsu_trap}(b)]. 

\section{\label{discussion}Discussion}
In this work, we analyzed pairing in two-dimensional Fermi gases above the superfluid 
transition temperature. For a gas composed of an equal number of spin-$\uparrow$ and 
spin-$\downarrow$ fermions, we found evidence of a crossover from a noninteracting 
gas of polarons to a pseudogap regime characterized by a BCS-like dispersion and reduced 
single-particle density of states near the Fermi energy. The details of this crossover as well 
as the properties of the effective polarons in a  balanced Fermi gas clearly call for further 
investigations.

We also analyzed two recent experiments that performed radio-frequency spectroscopy for 
2D Fermi gases~\cite{Sommer:2012,Zhang:2012}. To take into account the 
polaronic properties of the final-state atoms, we introduced a phenomenological model 
that uses dressed Green's functions for both initial- and final-state atoms in the calculation of the 
rf spectrum. Although this approach does not include the vertex 
corrections~\cite{Perali:2008, Pieri:2009,Pieri:2011}, our model qualitatively explains the different 
observations in Refs.~\cite{Sommer:2012,Zhang:2012}. In particular, our 
polaron-to-pseudogap crossover can be used to explain the apparent dichotomy in 
Refs.~\cite{Sommer:2012,Zhang:2012}. Our calculations suggest that Ref.~\cite{Sommer:2012} 
probed the pseudogap regime, while Ref.~\cite{Zhang:2012} provided data from the polaron 
side of the crossover. 

In Ref.~\cite{Sommer:2012}, the measured binding energies were found to be close to the 
energies of the corresponding two-body bound states. We demonstrated that this effect  
arises at the high-temperature regime, where many-body contributions to the experimentally 
measured thresholds in the rf spectrum become negligibly small. In order to clearly see the effect 
of many-body corrections to the measured binding energies, the temperature should be lower or 
the system should be quickly quenched to the strongly interacting regime as suggested in 
Ref.~\cite{Pietila:2012}.

Another interesting and theoretically largely unexplored direction is the dimensional 
crossover from 2D to 3D. In Ref.~\cite{Sommer:2012}, the dimensional crossover was explicitly 
probed, and in Ref.~\cite{Zhang:2012}, implications of the quasi-2D nature of the gas seem to be 
unavoidable since the Fermi energy was larger than the level splitting in the tightly confined 
direction. The formalism constructed for molecule formation in quasi-2D 
systems~\cite{Pietila:2012} has recently been extended to study the polaron 
problem~\cite{Levinsen:2012} and could be used to investigate the rf spectroscopy of quasi-2D 
systems.

{\it Note added in proof}: Recently, two preprints discussing the pseudogap phase in 2D Fermi gases have appeared~\cite{Klimin:2012,Watanabe:2012}.

\begin{acknowledgements}
We thank T.~Oka for discussions and S.~Kissel for proofreading the manuscript. This work 
was financially supported by the Finnish Cultural Foundation and Harvard-MIT CUA.

\end{acknowledgements}

\bibliography{manu}

\end{document}